\documentclass[12pt]{article}
\begin{document}
\pagenumbering{arabic}
\title{The angular momentum of plane-fronted gravitational waves in 
the teleparallel equivalent of general relativity}

\author{J. F. da Rocha-Neto$\,^{\dagger}$ and J. W. Maluf$\,^{\star}$}
\date{}
\maketitle

\begin{center}
Instituto de F\'isica, Universidade de Bras\'ilia, \\
70910-900, Bras\'ilia, DF, Brazil.
\end{center}

\begin{abstract}
We present a simplified expression for the gravitational angular 
momentum in the framework of the teleparallel equivalent of general 
relativity (TEGR). The expression arises from the constraints 
equations of the Hamiltonian formulation of the TEGR. 
We apply this expression to the calculation of the angular momentum 
of plane-fronted gravitational waves in an arbitrary 
three-dimensional  volume $V$ of space and compare the results
with those obtained for linearised gravitational waves. 
\end{abstract}


{\footnotesize
\noindent $\dagger$ rocha@fis.unb.br\\
\noindent $\star$ wadih@unb.br\\
\noindent $\star$ jwmaluf@gmail.com}

\bigskip
\section{Introduction}

\noindent

In the theory of general relativity, gravity is usually considered 
a phenomenon resulting from the curvature of space-time. The 
curvature of space-time is produced by the presence of matter, although there 
are vacuum solutions of the field equations. When massive objects move in
space-time, the curvature changes as a consequence of the movement of the 
objects. Massive objects that either acquire accelerated motion, or that undergo
unexpected changes, generate disturbances in space-time that travel away from 
the object. In Einstein's theory these disturbances generate gravitational 
radiation and gravitational waves \cite{Barish}. 
Although  gravitational waves have not yet been detected directly, in principle 
they should carry energy, momentum and angular 
momentum as gravitational radiation \cite{Dereli, Luke1, Luke2}. 
Nowadays it is clear that gravity can also be described by tetrad fields
and the torsion tensor, in the context of the teleparallel equivalent of
general relativity \cite{Maluf-AndP}. The accelerated motion of massive
bodies leads to changes in the space-time torsion as well.

The direct detection of gravitational waves will certainly 
open a new page in the history of general relativity. For example, by
means of these waves, the theory of general relativity can be tested
in the limit of strong gravitational fields, and not only as corrections
to the Newtonian theory or tests based on linearised solutions of
Einstein's equations. These waves carry important information
about their source. In addition,
the detection of these waves may provide insights about the
properties of the waves such as velocity,
helicity and polarization states\cite{Barish}. 

The detection of gravitational waves requires an appropriate 
description of physically observable quantities associated with them, such as
energy, momentum and angular momentum. 
As we know, in the realm of general relativity, 
there are no widely accepted definitions for these quantities, which are 
usually analysed by means of pseudotensors. Recently, in the context of 
linearised gravitational waves, the angular momentum of the waves has been 
investigated in the framework of pseudotensors \cite{Luke1}. However, an 
analysis of the gravitational angular momentum by means of pseudotensores 
is not satisfactory, because pseudotensores are coordinate 
dependent quantities. 

In this work, we first recall that in the framework of the teleparallel 
equivalent of general relativity, the problems of energy, momentum and angular 
momentum of gravitational waves are issues that are well established. 
The expressions for these quantities arise from the integral form of the
constraint equations of the Hamiltonian formulation of the TEGR. The 
resulting expressions are invariant under coordinate transformations of the 
three-dimensional space, and under time reparametrizations.   Here we apply 
the newly achieved, simplified definition of the gravitational angular momentum 
to plane-fronted gravitational  waves in an arbitrary three-dimensional volume 
$V$ of space. We also present the non-zero components of the  energy-momentum 
vector of the waves, and compare the results  obtained for plane-fronted 
gravitational waves with those obtained for linearised
gravitational waves. 

In order to check the consistency of the results obtained here, 
we analyze  the behaviour of a particle of mass $m$ in the presence
of gravitational waves discussed in this work, in order to analyse 
how the wave modifies the kinematic state of the particle by 
means of transfer of energy.

We use the following notation: spacetime indices $\mu, \nu, ...$
and $SO(3,1)$ indices $a, b, ...$ run from 0 to 3. Time and space
indices are indicated according to $\mu = 0, i$, $a = (0), (i)$. The
tetrad fields are indicated by $e^{a}\,_{\mu}$, and  the flat 
Minkowski spacetime metric tensor $\eta_{ab} = e_{a\mu}e_{b\nu}g^{\mu\nu}
= (-1, 1, 1, 1)$ raises and lowers tetrad indices.  The determinat of
tetrad fields is representad by $e = det(e^{a}\,_{\mu}) = \sqrt{-g}$
and we use the constants $G = c =1.$

\section{The Lagrangian and Hamiltonian formulations of the TEGR}

\noindent

In this section we  present a summary of both the Lagrangian 
and Hamiltonian formulations of the TEGR. For more details see 
Refs. \cite{Maluf-AndP,JF1,JF2,Hehl1, Hehl2, Ortin}.
The equivalence of the TEGR with Einstein's general relativity is obtained
by means of an indentity that relates the scalar curvature $R(e)$ constructed
out of the tetrad fields and a combination of quadratic terms in the
torsion tensor \cite{Maluf-AndP,JF1,JF2,Hehl1, Hehl2},

\begin{equation}
eR(e) \equiv -e\left({1\over 4}T^{abc}T_{abc} + 
{1\over 2}T^{abc}T_{bac} - T^{a}T_{a}\right)
+ 2\partial_{\mu}(eT^{\mu})\,,
\label{2.1}
\end{equation}
where $e = det(e^{a}\,_{\mu}), T_{a} = T^{b}\,_{ba}, 
T_{abc} = e_{b}\,^{\mu}e_{c}\,^{\nu}T_{a\mu\nu}$
and $T_{a\mu\nu} 
= \partial_{\mu}e_{a\nu} - \partial_{\nu}e_{a\mu}$ is the torsion tensor. 
The Lagrangian density in the TEGR is given by 
\begin{eqnarray}
L(e) &=& -e\left({1\over 4}T^{abc}T_{abc} + {1\over 2}T^{abc}T_{bac} 
- T^{a}T_{a}\right) - L_{M}\nonumber \\
& \equiv & -ke\Sigma^{abc}T_{abc} - L_{M}\,,
\label{2.2}
\end{eqnarray}
where $k = 1/(16\pi)$, $L_{M}$ represent the Lagrangian density for the matter 
fields, and $\Sigma^{abc}$ is defined by
\begin{equation}
\Sigma^{abc} = {1\over 4}\left(T^{abc} + T^{bac} - T^{cab}\right) 
+ {1\over 2}\left(\eta^{ac}T^{b} - \eta^{ab}T^{c}\right)\,.
\label{2.3}
\end{equation}
The quadratic combination, $\Sigma^{abc}T_{abc}$, except for a total divergence, 
is proportional to the scalar curvature $R(e)$. 

The variation of $ L(e)$ with respect to $e^{a\mu}$
yields the fields equations which, after some rearrangements, may be written 
in the form

\begin{equation}
\partial_{\nu}(e\Sigma^{a\mu\nu}) = 
{1\over 4k}ee^{a}\,_{\nu}(t^{\mu\nu} + T^{\mu\nu})\,,
\label{2.4a}
\end{equation}
where 
\begin{equation}
t^{\mu\nu} = k(4\Sigma^{bc\mu}T_{bc}\,^{\nu} - g^{\mu\nu}\Sigma^{bcd}T_{bcd})\,,
\label{2.4b}
\end{equation}
is interpreted as the gravitational energy-momentum tensor \cite{Maluf-AndP}  and 
$T^{\mu\nu} = e_{a}\,^{\mu}T^{a\nu}$, where 
$eT_{a\mu} = \delta L_{M}/\delta e^{a\mu}$. These field equations are 
equivalent to Einstein's equations. It
is possible to verify that the equations can be written as 
${1\over 2}(R_{a\mu}(e) - {1\over 2}e_{a\mu}R(e))$, provided we make $L_M=0$.

In order to obtain the Hamiltonian density in the TEGR we rewrite the Lagrangian
density $L$ in the phase space. To do this, first we note that the
Lagrangian density (\ref{2.2}) does not depend on the time derivative of the
tetrad component $e_{a0}$. Therefore this component  appears as a Lagrange
multiplier in the Hamiltonian density $H$.  From  (\ref{2.2}) we can
obtain the momentum canonically conjugated to $e_{ai}$ as 
$\Pi^{ai} = \delta L/\delta \dot{e}_{ai}= -4k\Sigma^{a0i}$, 
and $\Pi^{a0} \equiv 0$. The Hamiltonian density is obtained by rewriting the 
Lagrangian density in the form $L = \Pi^{ai}\dot{e}_{ai} - H$,
in terms of $e_{ai}, \Pi^{ai}$ and Lagrange multipliers. After the Legendre
transform is performed, we obtain the final form of the Hamiltonian density, that
is given by \cite{JF2}
\begin{equation}
H(e,\Pi) = e_{a0}C^{a} + \lambda_{ab}\Gamma^{ab}.
\label{2.5}
\end{equation}
In the above equation we have omitted a surface term. $C^{a} =
\delta H/\delta e_{a0}$  is a very long expression of the field variables,
and  $\Gamma^{ab}$ is defined by
 $\Gamma^{ab} = 2\Pi^{[ab]} + 4ke(\Sigma^{a0b} - \Sigma^{b0a})$. 
In $H$, $e_{a0}$ and $\lambda_{ab}$ are Lagrange multipliers which, after solving 
the field equations, are 
identified as $\lambda_{ab} = (1/4)(T_{a0b} - T_{b0a}+ e_{a}\,^{0}T_{00b} -
e_{b}\,^{0}T_{00a})$. The quantities $C^{a}$ and $\Gamma^{ab}$, as functions of
$\Pi^{ai}$ and $e_{ai}$, are first class constraints \cite{JF2}. 
It is possible to show that, in terms of the Poisson brackets, these constraints 
satisfy an algebra similar to the algebra of the Poincar\'e group \cite{JF2}.

The form of the constraints $C^{a}$ is given by
\begin{equation}
C^{a} = -\partial_{i}\Pi^{ai} + h^{a} = 0\,.
\label{2.9}
\end{equation}
Where $h^{a}$ is a intricate expression of the field quantities.

The relevant result presented in this work is that  the constraint 
$\Gamma^{ab}$ can be simplified and rewritten as a total divergence
according to
\begin{equation}
\Gamma^{ab} = 2\Pi^{[ab]} - 2k\partial_{i}[e(e^{ai}e^{b0} - e^{bi}e^{a0})] = 0\,. 
\label{2.10}
\end{equation}
Two important results of the TEGR are the following: both constraints $C^a$ and 
$\Gamma^{ab}$ may be written in terms of a total divergence which, under
integration, yield straightforward expressions for the gravitational 
energy-momentum and angular momentum,
respectively.  In particular, the integral form of the
constraint $C^{a} = 0$ yields the gravitational energy-momentum 
$P^{a}$ \cite{JF3},
\begin{equation}
P^{a} = - \int_{V}d^{3}x\partial_{i}\Pi^{ai}\,,
\label{2.11}
\end{equation}
where $V$ is an arbitrary volume of the three-dimensional space and 
$\Pi^{ai} = -4k\Sigma^{a0i}$.

In Ref. \cite{JF3}, motivated by the definition of $P^{a}$ as an integral of a
total divergence, we presented an expression for the angular momentum of the 
gravitational field out of the integral form of the constraint 
$\Gamma^{ij} = 2\Pi^{[ij]} = e_{a}\,^{i} \Gamma^{ab}e_{b}\,^{j} = 0$, where
$2\Pi^{[ij]}$ was taken as the gravitational angular momentum density. 
In Ref. \cite{JF4}, by writing the Hamiltonian $H$ in a more simple form, we 
redefined the angular momentum of the gravitational field as the integral of 
the constraint $\Gamma^{ab} = 0$. Unlike the expressions presented in 
\cite{JF3, JF4}, here we note that, after some simplifications, the constraint
$\Gamma^{ab} = 0$ given by Eq. (\ref{2.10}) may be written in terms of a total 
divergence. 

In similarity with the definition given in \cite{JF4}, we define the 4-angular
momentum of the gravitational field as
\begin{equation}
L^{ab} = -\int_{V}d^{3}xM^{ab}\,,
\label{2.12}
\end{equation}
where 
\begin{equation}
M^{ab} =  2\Pi^{[ab]} =(\Pi^{ab} - \Pi^{ba}) 
= 2k\partial_{i}[e(e^{ai}e^{b0} - e^{bi}e^{a0})]\,.
\label{2.13}
\end{equation}
Expressions  (\ref{2.11}) and (\ref{2.12}) are both invariants under coordinate
transformations of the three-dimensional space, and under time reparametrizations.
Note that the integrals in (\ref{2.11}) and (\ref{2.12}) may be carried out on 
surfaces that enclose an arbitrary volume $V$. The expression above generalizes
the one obtained in \cite{Maluf-Ulhoa}. The latter was also given as a total
divergence, but the result was obtained for the metric tensor with axial
symmetry..

It is possible to show, using Poisson brackets, that the quantities $P^{a}$ and 
$L^{ab}$ defined in (\ref{2.11})
and (\ref{2.12}) respectively, satisfy the algebra of the Poincar\'e group,
\begin{equation}
\{P^{a}, P^{b}\} = 0\,,
\label{2.14}
\end{equation}
\begin{equation}
\{P^{a},L^{bc}\} = \eta^{ab}P^{c} - \eta^{ac}P^{b} \,,
\label{2.15}
\end{equation}
\begin{equation}
\{L^{ab},L^{cd}\} = \eta^{ad}L^{cb} + \eta^{bd}L^{ac} - 
\eta^{ac}L^{db} - \eta^{bc}L^{ad}  \,.
\label{2.16}
\end{equation}
Therefore, from a physical point of view, the interpretation of the 
quantities $P^{a}$ and $L^{ab}$ is consistent.

In the next section we will apply the definition of $L ^{ab}$ to the
calculation of the components of the angular momentum carried by a  
plane-fronted gravitational wave. We will also present the non-zero components 
of the energy-momentum vector calculated in Ref. \cite{Maluf1}, and 
compare these results with those obtained for linearised gravitational waves.

\section{The angular momentum of gravitational waves}

\noindent

A plane-fronted gravitational wave propagating in the $z$ direction
may be described by the line element \cite{Kramer}
\begin{equation}
ds^{2} = dx^{2} + dy^{2} + 2dudv + H(x,y,u)du^{2}\,.
\label{2.17}
\end{equation}
The vacuum field equations are reduced to
\begin{equation}
\left({\partial^{2}\over \partial x^{2}} + 
{\partial^{2}\over \partial y^{2}}\right) H(x,y,u) = 0\,.
\label{2.18}
\end{equation}
By writing the coordinates $(u,v)$ in terms of the coordinates $(t,z)$, 
$$ u = {1\over \sqrt{2}}(z - t)\,, \;\;\;\;\;\; v = 
{1\over \sqrt{2}}(z + t),$$
the line element in (\ref{2.17}) becomes 
\begin{equation}
ds^{2} = \left({H\over 2} - 1\right) dt^{2} + dx^{2} + dy^{2} +
\left({H\over 2}+1\right) dz^{2}
- H dt dz\,.
\label{2.19}
\end{equation}
In the following calculations  we need the inverse of the metric tensor. 
It is given by
\begin{equation}
g^{\mu\nu} = \left(\begin{array}{cccc}
-1 - {H\over 2} & 0 & 0 & -{H\over 2}\\
0 & 1 & 0 & 0\\
0 & 0 & 1 & 0\\
-{H\over 2} & 0 & 0 & 1- {H\over 2}
\label{2.20}
\end{array}
\right).
\end{equation}

In order to evaluate  $L^{ab}$ associated with the gravitational wave 
described by (\ref{2.19}), we  consider a set of tetrad fields adapted to 
static observers. These tetrad fields must satisfy $e_{(0)}\,^{i} = 0$, and 
are given by

\begin{equation}
e_{a\mu} = \left(\begin{array}{cccc}
-A & 0 & 0 & -B\\
0 & 1 & 0 & 0 \\
0 & 0 & 1 & 0\\
0 & 0 & 0  & C
\end{array}
\right),
\label{2.21}
\end{equation}
where 

$$A = \left(1 - {H\over 2}\right)^{1/2}\,, \;\;\;\;\;  
AB = {H\over 2} \,,\;\;\;\;\;   AC = 1\,. $$

The tetrads in (\ref{2.21}) are adapted to static observers. They are better 
understood if we consider the inverse  $e_{a}\,^{\mu}$. In terms of 
components, we have
\begin{equation}
e_{(0)}\,^{\mu} = (1/A, 0, 0, 0),
\label{2.22}
\end{equation}
and 
\begin{equation}
e_{(1)}\,^{\mu} = (0, 1, 0, 0),\;\;\;\;\;  
e_{(2)}\,^{\mu} = (0, 0, 1, 0), \;\;\;\;\;  e_{(3)}\,^{\mu} = (-H/(2A), 0, 0, A).
\label{2.23}
\end{equation}
The four velocity of the frame is given by $u^{\mu} = e_{(0)}\,^{\mu}$. Therefore,
equation (\ref{2.22}) fixes the kinematic state of the frame:  the three velocity
conditions $e_{(0)}\,^{i} = 0$, ensure that the frame is static.  The other three
conditions in (\ref{2.23}) fix the spatial orientation of the frame, i.e., 
$e_{(1)}\,^{\mu},  e_{(2)}\,^{\mu}$  and  $e_{(3)}\,^{\mu}$
are  unit vectors along the $x, y$ and $z$ directions, respectively. 
An alternative way to characterize the
tetrad fields is by fixing the six components of the acceleration tensor
$\phi_{ab} = -\phi_{ba}=  (1/2)(T_{(0)ab} + T_{a(0)b} - T_{b(0)a})$
\cite{Maluf2}, where $\phi_{a(0)}$ and $\phi_{(i)(j)}$ are
translational and rotational accelerations, respectively. These accelerations are 
necessary to maintain the frame in
a given inertial state in spacetime. Here we note that if we take 
$H = 0$, $e_{a}\,^{\mu} = \delta_{a}^{\mu},$
and therefore $T_{a\mu\nu} = 0$. 

Although the components of $M^{ab}$ do not depend on the torsion tensor 
explicitly, the latter is important to calculate the energy-momentum 
vector $P^{a}$ and the acceleration tensor $\phi_{ab}$.  For the tetrad fields 
given by (\ref{2.21}), the non-vanishing components $T_{\mu\nu\lambda}$ were
calculated in Ref. \cite{GRG2013}. In the latter reference we addressed the
non-linear gravitational wave given by Eq. (\ref{2.17}), and the same tetrad 
fields described by Eq. (\ref{2.21}). We showed that the torsion tensor obtained
in the framework of these gravitational waves breaks parallelograms in space-time,
and obtained a physical result that cannot be achieved in the realm of the 
standard formulation of general relativity. We will return to this point in the
final remarks of this article.

In order to calculate the components of the angular momentum transported  by the 
waves described by (\ref{2.19}), we need the quantities given by (\ref{2.22}) and
(\ref{2.23}). The components of the angular momentum  can be  obtained as 
surface integrals. However, since we do not dispose of the explicit form of $H$,
we find it more appropriate to calculate these components as volume integrals, 
according to 
\begin{equation}
L^{ab} = -2k\int_{V}\partial_{j}[e(e^{aj}e^{b0} - e^{a0}e^{bj})]d^{3}x.
\label{2.25}
\end{equation}
The determinant  of tetrad fields is $e = AC = 1$. From (\ref{2.22}), (\ref{2.23})
and (\ref{2.25}) we obtain the nonvanishing components of $L^{ab}$. They read 
\begin{equation}
L^{(0)(1)} = -2k\int_{V}\partial_{x}\left(1\over A\right)d^{3}x, 
\label{28a}
\end{equation}
\begin{equation}
L^{(0)(2)} = -2k\int_{V}\partial_{y}\left(1\over A\right)d^{3}x, 
\label{28b}
\end{equation}
\begin{equation}
L^{(1)(3)} = 2k\int_{V}\partial_{x}\left(H\over 2A\right)d^{3}x, 
\label{28c}
\end{equation}
\begin{equation}
L^{(2)(3)} = 2k\int_{V}\partial_{y}\left(H\over 2A\right)d^{3}x.  
\label{28d}
\end{equation}
We observe that the plane-fronted gravitational waves do not carry angular 
momentum in the direction of  propagation, i.e., $L^{z} = L^{(1)(2)} = 0$,
but only in the orthogonal direction.

The non-vanishing components of the gravitational energy-momentum vector 
$P^{a}$ of plane-fronted gravitational waves have been calculated in Ref. 
\cite{Maluf1}. However, we present them here to compare their values with
those obtained for linearised gravitational waves. The non-zero expressions
of $P^{a}$ given by (\ref{2.11}), constructed out of Eq. (\ref{2.21}), read

\begin{equation}
P^{(0)}= P^{(3)} = -{k\over 8}\int_{V} \left[{1\over A^{3}}\lbrack (\partial_{x}H)^2+(\partial_{y}H)^{2}\rbrack \right]d^{3}x.
\label{2.26a}
\end{equation}
We note that $P^{2} = \eta_{ab}P^{a}P^{b} = 0$. This result is consistent 
with the fact that the gravitational waves should describe massless particles.
We will now compare this result with that obtained in the context of a 
linearised gravitational wave propagating in the $z$ direction. 

In the diagonal polarization, the linearised gravitational wave is described 
by the line element,

\begin{equation}
ds^2 = -dt^2 + [1+f(t-z)]dx^2 + [1 - f(t-z)]dy^2 +dz^2\,,
\label{2.27}
\end{equation}
where $|f(t-z)| << 1$ and $e = \sqrt{-g} = 1$.  
The tetrad fields adapted to static observers, in the space-time determined by 
(\ref{2.27}), is given by

\begin{equation}
e_{a\mu} = \left(\begin{array}{cccc}
-1 & 0 & 0 & 0\\
0 & 1+{f\over 2} & 0 & 0 \\
0 & 0 & 1-{f\over 2} & 0\\
0 & 0 & 0 & 1
\end{array}
\right)\,.
\label{2.28}
\end{equation}
The non-vanishing components of the torsion tensor $T_{a\mu\nu}$  are

\begin{equation}
T_{(1)01} ={1\over 2} \partial_{t}f,\,\,\,\, T_{(2)02} = 
-{1\over 2}\partial_{t}f, \,\,\,\, T_{(1)13} = 
-{1\over 2}\partial_{z}f\,,\,\,\,\, T_{(2)23} = {1\over 2}\partial_{z}f\,.
\label{2.29}
\end{equation}
We notice that if we consider only first order tems in $f$, all 
components of the acceleration tensor  $\phi_{ab}$ vanish. This fact 
means, as we will see in the next section, that linearised gravitational 
waves do not transfer energy to a particle of mass $m$. Since the inverse
of the tetrad fields above is also represented by a diagonal matrix, it is not
difficult to conclude from Eq. (\ref{2.25}) that the components  of the angular
momentum carried by linearised gravitational waves are given by 

\begin{equation}
L^{(0)(i)} = 2k\int_{V}\partial_{j}(ee^{(i)j})d^3x = 0,\,\,\,\,\,\, L^{(i)(k)} = 0,
\label{2.31}
\end{equation}
where in the first equation above we use the fact that $e = 1$, and 
$e^{(1)1} = 1 -f/2$ and $e^{(2)2} = 1+f/2$ are functions only of $t-z$. This 
result implies that in the framework of the TEGR, gravitational waves in the
linearised approximation do not transport angular momentum. In the linear
approximation, and in the context of the tetrad fields (\ref{2.28}), we can 
easily show that all the components of $\Sigma^{a0j}$ vanish. Therefore, all 
components of $P^{a}$ given by Eq. (\ref{2.11}) are also zero. Thus, unlike 
plane-fronted gravitational waves, gravitational waves in the linearised 
approximation do not carry energy, momentum and angular momentum.

\section{Polarization of plane-fronted gravitational waves}

The polarization of plane-fronted gravitational waves may be obtained from 
the analysis of the acceleration tensor $\phi_{ab}$, and the behaviour of a 
free particle in the gravitational field. We recall that $\phi_{ab}$
represent the inertial (i.e., non-gravitational) accelerations that are 
necessary to maintain the frame in a given inertial state. In the
present context, we are considering a static frame in space-time, defined by
Eqs. (\ref{2.22}) and (\ref{2.23}). The values of $\phi_{ab}$ that follow
from Eq. (\ref{2.21}) and the latter equations are given by 

\begin{equation}
\phi_{(0)(1)} = -{1\over 4A^{2}}\partial_{x}H,
\label{p-01}
\end{equation}

\begin{equation}
\phi_{(0)(2)} = -{1\over 4A^{2}}\partial_{y}H,
\label{p-02}
\end{equation}

\begin{equation}
\phi_{(0)(3)} = {1\over 4A^{3}}\partial_{z}H,
\label{p-03}
\end{equation}

\begin{equation}
\phi_{(1)(2)} = 0,
\label{p-12}
\end{equation}

\begin{equation}
\phi_{(1)(3)} = -{1\over 4A^{2}}\partial_{x}H,
\label{p-13}
\end{equation}

\begin{equation}
\phi_{(2)(3)} = -{1\over 4A^{2}}\partial_{y}H.
\label{p-23}
\end{equation}

If we imagine a free particle in space-time that is {\it not} subject to
the inertial conditions established by Eqs. 
(\ref{2.21}-\ref{2.23}), then the presence of the wave will impart the 
{\it negative} values of the accelerations above on the particle. Equations
(\ref{p-01}) and (\ref{p-02}) imply that the wave is transversal, since the
particle will move in the transversal $x$ and $y$ directions (and possibly 
oscillate, depending on the form of the function $H$). Equation (\ref{p-03})
implies a longitudinal motion. The resulting motion of the free particle is,
to some extent, similar to the behaviour of an electron under the action of
an intense laser beam, in which case we know that the motion of the electron
results in the well known ``figure 8'', that generates the Thomson scattering.
Therefore, the plane-fronted gravitational wave is both a transversal and 
longitudinal wave.

\section{Massive particle in the presence of gravitational waves}
\noindent

In order to verify the consistency of the results obtained up to here, let us 
analyse the transfer of energy to a particle of mass $m$ in the presence of  
gravitational waves. First, from the point of view of the
Euler-Lagrange equations, we will consider the particle in the 
presence of a plane-fronted gravitational wave,
and then, in a similar way, we will consider the particle in the presence of a
linearised gravitational wave. 

For the particle in the presence of a plane-fronted gravitational wave described by
(\ref{2.19}), the Lagrangian $L ={m\over 2} g_{\mu\nu}\dot{x}^{\mu}\dot{x}^{\nu}$  
that describes the particle motion is written as

\begin{equation}
L ={m\over 2} \left[(H/2 - 1)\dot{t^{2}} + \dot{x}^{2} + \dot{y}^{2} + 
(H/2 + 1)\dot{z}^{2} - H\dot{t}\dot{z}\right], 
\label{2.32}
\end{equation}
where $\dot{x}^{\mu} = dx^{\mu}/d\tau$, with $\tau$ being the proper time of the 
particle. The Euler-Lagrange equations for the $t, x, y, z$ coordinates are,
respectively

\begin{equation}
2\ddot{t} + \sqrt{2}H\ddot{u} + \sqrt{2}\dot{H}\dot{u} 
- {1\over \sqrt{2}}{\partial H\over \partial u}\dot{u}^2 = 0,
\label{34a}
\end{equation}
\begin{equation}
2\ddot{x} - {\partial H\over \partial u}\dot{u}^{2} = 0,
\label{34b}
\end{equation}
\begin{equation}
2\ddot{y} - {\partial H\over \partial u}\dot{u}^{2} =  0,
\label{34c}
\end{equation}
\begin{equation}
2\ddot{z} +  \sqrt{2}H\ddot{u} + \sqrt{2}\dot{H}\dot{u} 
- {1\over \sqrt{2}}{\partial H\over \partial u}\dot{u}^2 = 0.
\label{34d}
\end{equation}
From the first and fourth equations above we get

\begin{equation}
\ddot{z} - \ddot{t} = 0 \to \ddot{u} = 0 \to \dot{u}=
{1\over \sqrt{2}}(\dot{t} - \dot{z}) \equiv k = constant.
\label{2.34}
\end{equation}

We recall that the tetrad fields (\ref{2.21}) are interpreted as a reference
frame adapted to static observers in space-time, since we identify 
the $e_{(0)}\,^{\mu}$ components of the frame with the 
four-velocities $u^{\mu}$ of the observers, 
$e_{(0)}\,^{\mu} = u^{\mu} = (1/A, 0, 0, 0)$.
Thus, if we consider the particle of mass $m$ in the presence of plane-fronted 
gravitational waves, its four-momentum, when measured by an observer adapted 
to (\ref{2.21}), will be given by $p_{a} = e_{a}\,^{\mu}p_{\mu}$,  where
$p^{\mu} = m\dot{x}^{\mu}$. Therefore, the energy of the particle 
measured by observers adapted to (\ref{2.21}) is given by

\begin{equation}
E = - p_{\mu}u^{\mu} =-e_{(0)}\,^\mu p_\mu
= - e_{(0)}\,^{0}p_{0} = - {p_{0}\over A}\,,
\label{2.35}
\end{equation}
where $p_{0} = g_{0\mu}p^{\mu} = g_{00}m\dot{t} + g_{03}m\dot{z}.$ Substituting
$p_{0}$ in the equation above and expressing $A$, $g_{00}$ and $g_{03}$ in terms
of $H$, we are left with
\begin{equation}
E = m[(1- H/2)^{1/2}\dot{t} + (H/2)(1-H/2)^{-1/2}\dot{z}]\,.
\label{2.36}
\end{equation}
From the line element (\ref{2.19}) it follows that  
\begin{equation}
d\tau^{2} = (-g_{00})dt^{2} \to \dot{t} = {dt\over d\tau}= 
\left(1 - {H\over 2}\right)^{-1/2}\,.
\label{2.37}
\end{equation}
Using (\ref{2.36}) and the equation above, we can eliminate $\dot{z}$ and 
$\dot{t}$ in terms of the constant $k$ and the function $H$, so that we may 
rewrite  $E$  as
\begin{equation}
E = m\left(1 + {H/2\over 1- H/2}\right) + 
m\sqrt{2}k\left({H/2\over 1- H/2}\right)\,.
\label{2.38}
\end{equation}
Here we note  that the plane-fronted gravitational wave changes the 
kinematic state of the particle by means of energy transfer.  In particular,
if we take $k = 0$, and if $H < 0$, the energy of the particle in the presence
of the wave will be lower than its rest energy, i. e. $E < E_{0} = m$.

Let us now consider the particle  of mass $m$ in the presence of a linearised 
gravitational wave.  For  the line element in (\ref{2.27}), the  Lagrangian is
\begin{equation}
L = {m\over 2}\left(-\dot{t}^{2}+g_{11}\dot{x}^{2}+
g_{22}\dot{y}^{2}+\dot{z}^{2}\right)\,,
\label{2.39}
\end{equation}
and the Euler-Lagrange equations in the coordinates $t, x, y, z$ reduce to
\begin{equation}
2\ddot{t}  + {\partial g_{11} \over \partial u}\dot{x}^{2} 
+ {\partial g_{22}\over \partial u}\dot{y}^{2} = 0,
\label{40a}
\end{equation}
\begin{equation}
{d\over d\tau}(g_{11}\dot{x}) = 0,
\label{40b}
\end{equation}
\begin{equation}
{d\over d\tau}(g_{22}\dot{y}) = 0,
\label{40c}
\end{equation}
\begin{equation}
2\ddot{z}  + {\partial g_{11} \over \partial u}\dot{x}^{2} 
+ {\partial g_{22}\over \partial u}\dot{y}^{2} = 0,
\label{40d}
\end{equation}
Again, from the first and fourth equations above we have $\ddot{t} 
- \ddot{z} = 0 \to \dot{t} - \dot{z} = k' = constant$.

The four-velocity $u^{\mu}$ of observers adapted to the tetrad fields given by
(\ref{2.28}) is $u^{\mu} = e_{(0)}\,^{\mu} = (1, 0, 0, 0.)$  Therefore, 
the energy of the particle in the presence of a linearised gravitational wave,
measured by observers adapted to (\ref{2.28}), is 
$E = - p_{\mu}u^{\mu} = - p_{0} =  p^{0} = m(dt /d\tau)$. Since $dt/ d\tau  
= 1/(-g_{00})^{1/2} = 1$, the energy of the particle in the presence
of a linearised gravitational wave is simply 
\begin{equation}
E=m\,. 
\label{2.41}
\end{equation}
So, from the point of view of static observers, the linearised gravitational 
wave does not transfer energy to the particle. This result is consistent with
the previously obtained conclusion in Section 3, namely, that linearised 
gravitational waves do not transport energy. 

\section{Final remarks}

\noindent

In this work we have presented a very simple expression for the density of 
angular momentum of the gravitational field. The expression, given by Eq. 
(\ref{2.13}), arises in the constraint equations of the Hamiltonian 
formulation of the TEGR and is given by a total divergence. 
The definition in question is invariant under coordinate transformations of the
three-dimensional space, and under time reparametrizations. In addition, the 
definitions of energy-momentum and angular momentum presented in equations 
(\ref{2.11}) and (\ref{2.12}) are both invariant under global S0(3,1) 
transformations of the frame. However, we must remember that in the 
theory of special relativity, physical quantities such as energy and momentum
are frame dependent, and  we believe that a similar property is expected to hold
in the theory of general  relativity. Moreover, in view of Eq. (\ref{2.12}),
we see that the gravitational angular momentum enclosed by a 
three-dimensional volume $V$ can be calculated for arbitrary values of the 
volume $V$.

We applied the definition given by Eq. (\ref{2.12}), and calculated the  
components of the angular momentum  within a volume $V$, carried by (i) a 
plane-fronted gravitational wave, and by (ii) a linearised gravitational wave. 
All calculations were performed  with respect to an observer adapted to a static
reference frame. Unlike the case for plane-fronted gravitational waves, 
the angular momentum carried by linearised gravitational wave vanishes. For the 
plane-fronted gravitational wave, the component of the angular momentum in the
direction of propagation of the wave vanishes. From the point
of view of the TEGR, we show that, contrary to what happens in the case of 
plane-fronted gravitational waves, linearised gravitational waves do not carry
neither energy nor angular momentum.

In order to better understand the results obtained in the present context, we have 
reviewed in section 5 the change of the energy of a particle of mass $m$, in a 
frame adapted to static observers, in the presence of gravitational waves.
Equation (\ref{2.38}) shows that in the presence 
of a plane-fronted gravitational wave, the energy of the particle is modified, and
energy transfer may occur between the particle and the wave. Therefore, we conclude 
that in this case the wave carries energy. When we consider the particle
in the presence of a linearised gravitational wave, we see from Eq. (\ref{2.41}) 
that the energy of the particle adapted to a static reference frame is
not modified. Therefore we conclude that linearised gravitational waves do not 
transfer energy to the particle.

The equivalence of the TEGR with the standard formulation of general relativity
holds at the level of field equations. However, in the TEGR there are physical
predictions that cannot be achieved in the standard, metrical formulation of the
theory. In Ref. \cite{GRG2013} we have considered an experimental setup similar
to the two perpendicular arms of the interferometer constructed at the LIGO 
experiment. Likewise, we assumed that a laser beam travels back 
and forth along the two arms, at the end of which there are {\it fixed} mirrors.
Considering that the two perpendicular arms are established in the $xy$ plane,
say, we found that in the presence of a plane fronted gravitational wave that 
travels in the $z$ direction there is a lapse of time between the departure 
(at the same point in the three-dimensional space) and arrival of the two laser 
beams, along the perpendicular directions. This lapse of time is a consequence 
of the breaking of a certain parallelogram in space-time \cite{GRG2013}, and 
cannot be obtained in the realm of the metrical formulation of
general relativity, since it is described by the space-time torsion tensor.
This is an interesting, distinctive feature that (i) cannot be established in 
the context of linearised gravitational waves, (ii) is feasible of being 
measured, and (iii) that cannot be explained in terms of the curvature tensor.
In the TEGR we deal with tetrad fields, which, in a certain sense,
are the ``square root'' of metric tensor. Thus, it is natural that some physical
predictions of the TEGR do not have the corresponding counterpart in the
metrical formulation.

\end{document}